\documentclass[prl,twocolumn,floatfix,nobibnotes,superscriptaddress]{revtex4}
\usepackage{amsmath}
\usepackage{graphicx}
\usepackage[tight]{subfigure}

\usepackage[usenames,dvipsnames]{color}

\newcommand{\Tr}{\ensuremath \mathrm{Tr}}
\usepackage{color}

\begin{document}
\title{Interatomic exchange interactions in non-collinear magnets}
\author{A. Szilva}
\affiliation{Department of Physics and Astronomy, Division of Materials Theory, Uppsala University,
Box 516, SE-75120, Uppsala, Sweden}
\author{M. Costa}
\affiliation{Department of Physics and Astronomy, Division of Materials Theory, Uppsala University,
Box 516, SE-75120, Uppsala, Sweden}
\affiliation{Instituto de F\'isica, Universidade Federal Fluminense, 24210-346 Niter\'oi, Rio de Janeiro, Brazil}
\affiliation{Department of Physics and Astronomy, University of California, Irvine, California 92697, USA}
\author{A. Bergman}
\affiliation{Department of Physics and Astronomy, Division of Materials Theory, Uppsala University,
Box 516, SE-75120, Uppsala, Sweden}
\author{L. Szunyogh}
\affiliation{Department of Theoretical Physics and Condensed Matter Research Group of Hungarian Academy of Sciences, Budapest University of Technology and Economics, Budafoki \'ut 8. H1111 Budapest, Hungary}
\author{L. Nordstr\"om}
\affiliation{Department of Physics and Astronomy, Division of Materials Theory, Uppsala University,
Box 516, SE-75120, Uppsala, Sweden}
\author{O. Eriksson}
\affiliation{Department of Physics and Astronomy, Division of Materials Theory, Uppsala University,
Box 516, SE-75120, Uppsala, Sweden}

\begin{abstract}

We derive ab inito exchange parameters for general non-collinear magnetic configurations, in terms of a multiple scattering formalism. We show that the general exchange formula has an anisotropic-like term even in the absence of spin-orbit coupling, and that this term is large, for instance for collinear configuration in bcc Fe, whereas for fcc Ni it is quite small. We demonstrate that keeping this term leads to that one should consider a biquadratic effective spin Hamiltonian even in case of collinear arrangement. In non-collinear systems this term results in new tensor elements, that are important for exchange interactions at finite temperatures but they have less importance at low temperature. To illustrate our results in practice, we calculate for bcc Fe magnon spectra obtained from configuration dependent exchange parameters, where the configurations are determined by finite temperature effects. Our theory results in the same quantitative results as the finite temperature neutron scattering experiments.
  
\end{abstract}
\pacs{later}

\maketitle

The non-collinear magnetic alignment, when a global magnetization axis is not easily identifined, can appear as the ground state of  several magnetic materials \cite{kubler}, e.g. spin-spiral or spin-glass systems,  and for non-equilibrium or dynamical systems (e.g. $T>0$) it is  universal. Considering the rigid spin and adiabatic approximation,  the calculation of exchange interaction between well-defined moments is crucial for atomistic, first principle spin dynamics simulations and for the interpretations of experimental results. Although the formula in case  of collinear arrangement is known for a long time, due to the seminal work of Ref.~\cite{oldlicht}, even for correlated systems \cite{newlicht}, a counterpart for non-collinear arrangement is lacking. In this paper we  derive a general formula for both collinear and non-collinear spin-systems, where we make use of  the magnetic force theorem \cite{mac}, \cite{meth}. Furthermore, our analysis is expressed in terms of multiple scattering formalism (MSF) \cite{gyorffy}. Analyzing the one- and two-site spin rotations, we map the analytically derived exchange parameters onto effective spin Hamiltonians, and discuss their appropriateness for collinear and non-collinear spin arrangements, and we illustrate our results with a numerical simulation of the magnon energies of bcc Fe at elevated temperature.

The fundamental equation of a scalar relativistic MSF is given as \cite{gyorffy}%
\begin{equation}
\left( \tau _{ij}^{-1}\right) _{L\sigma ,L^{\prime }\sigma ^{\prime
}}=P_{iL\sigma \sigma ^{\prime }}\delta _{ij}\delta _{LL^{\prime
}}-G^{0}_{ij,LL^{\prime }}\delta _{\sigma \sigma ^{\prime }}\;,
\label{FEMSTT}
\end{equation}%
where $\tau _{ij}^{-1}$ stands for the scattering path operator (SPO), $P_{i}$ denotes the inverse of the single site scattering operator (ISO), $L$ stands for the angular momentum and magnetic quantum numbers, $\sigma$ refers to the spin-index, $G^{0}$ is the free (or bare) structure constant and indices $i$ and $j$ refer to the considered lattice sites. Later on in our presentation, we omit the orbital and spin indices. We introduce a general notation for the ISO as follows,%
\begin{equation}
P_{i}(\varepsilon )=p_{i}^{0}I_{2}+\vec{p}_{i}\vec{\sigma}%
=p_{i}^{0}I_{2}+p_{i}\vec{n}_{i}\vec{\sigma}\;,  \label{pii}
\end{equation}%
where the unit vector $\vec{n}_{i}$ refers to the magnetic spin moment at site $i$, $\vec{\sigma}$ is the Pauli-matrices, $I_{2}$ is the unit matrix in spin space, $p_{i}^{0}$ denotes the non-magnetic and the vector $\vec{p}_{i}$ stands for the magnetic part of the ISO. Introducing a similar notation for the SPO it can be written that%
\begin{equation}
\tau _{ij}(\varepsilon )=T_{ij}^{0}I_{2}+\vec{T}_{ij}\vec{\sigma}\;,
\label{tijj}
\end{equation}%
where the vector $\vec{T}_{ij}$ has three ($x$, $y$ and $z$) matrix components, which enables to treat a non-collinear arrangement. We write the variation of ISO as%
\begin{equation}
\delta P_{i}=p_{i}\delta \vec{n}_{i}\vec{\sigma}\;,  \label{deltapii}
\end{equation}%
where $\delta\vec{n}_{i}$ stands for the deviation of a spin moment after an infinitesimal rotation at site $i$. We also introduce the tensor
\begin{equation}
A_{ij}^{\alpha \beta }=\frac{1}{\pi }\int\limits_{-\infty }^{\varepsilon
_{F}}d\varepsilon \operatorname{Im}\Tr_{L}\left( p_{i}T_{ij}^{\alpha
}p_{j}T_{ji}^{\beta }\right) \;,  \label{Adeff}
\end{equation}%
where indices $\alpha $ and $\beta $ run over $0$, $x$, $y$ or $z$. We note that $A_{ij}^{\alpha \beta }=A_{ji}^{\beta \alpha }$ because of the properties of the trace. The collinear alignment is an important special case when the global coordinate system can be chosen so that the vector $\vec{T}_{ij} $ has only non-zero values of the $z$ component between every site, implying all $A_{ij}^{\alpha \beta}$-s are equal zero except for $A_{ij}^{00}$ and  $A_{ij}^{zz}$. We henceforth refer to $A_{ij}^{00}$ and  $A_{ij}^{zz}$ as the collinear exchange parameters, and the other elements as non-collinear exchange parameters. Introducing quantities $T_{ij}^{\uparrow }=T_{ij}^{0}+T_{ij}^{z}$ and $T_{ij}^{\downarrow }=T_{ij}^{0}-T_{ij}^{z} $ for collinear systems and using the time reversal symmetry,  one obtains the well-known expression  \cite{oldlicht}
\begin{equation}
A_{ij}^{00}-A_{ij}^{zz}=\frac{1}{\pi }\int\limits_{-\infty }^{\varepsilon
_{F}}d\varepsilon \operatorname{Im}\Tr_{L}\left( p_{i}T_{ij}^{\uparrow
}p_{j}T_{ji}^{\downarrow }\right) \; , \label{lichtdef0}
\end{equation}%
which can be defined for any (non-collinear) configuration as\begin{equation}
J_{ij}^\mathrm{L}=A_{ij}^{00}-A_{ij}^{zz}, \label{lichtdef}
\end{equation}%
and will be referred to as the LKAG-formula.

According to Andersen's local force theorem \cite{mac}, \cite{meth} the total energy variation can be written as the variation of the integrated density of states time energy (first moment). The so-called Lloyd formula \cite{lloyd} says how to calculate this from ISO and SPO, see Eqs. (\ref{pii}) and (\ref{tijj}), in the presence of any perturbation, e.g. for a rotation of a spin-moment. We speak of $one$-site spin rotation when this perturbation is given due to a rotated magnetic moment only at one site $i$ with infinitesimal angle $\delta \theta$. In that case the detailed derivation of the total energy variation is written in Appendix A, here we give the final result in collinear limit ($T_{ij}^{x}=T_{ij}^{y} \simeq 0$),
\begin{equation}
\delta E^\mathrm{one}_{i} =-2\sum\limits_{j\neq i}J_{ij}^\mathrm{L} \delta {n}_{i}^{z}  \;. \label{onelicht}
\end{equation}%
Note that this equation is obtained without making any assumption of an effective spin-Hamiltonian, and is instead a direct consequence of multiple scattering theory. Therefore any effective spin-Hamiltonian should reproduce the results of Eq. (\ref{onelicht}), in the collinear limit, as regards the energy of one-site rotations. It should be noted that $\delta{n}_{i}^{z}$ is proportional to $\left(\delta \theta \right)^{2}$, therefore $\sum_{j \neq i}J_{ij}^\mathrm{L}$ is positive for a ferromagnetic ground state. The effective (Weiss) field can be obtained from Eq. (\ref{onelicht}) and any desired spin Hamiltonian should recover it in the collinear limit.

Next, we consider $two$-site spin moment rotations, i.e., two spin moments at site $i$ and $j$ are rotated simultaneously in opposite directions with angle $\delta \theta$. Using the Lloyd formula we find  an interaction term appears  in the variation of the total energy expression. This variation can be written for the general, non-collinear case as
\begin{equation}
\delta E_{ij}^\mathrm{two}=\delta E_{ij}^\mathrm{HT}+\delta E_{ij}^\mathrm{AT}\;,  \label{E1}
\end{equation}%
where%
\begin{equation}
\delta E_{ij}^\mathrm{HT}=-2\left( A_{ij}^{00}-\sum_{\mu =x,y,z}A_{ij}^{\mu \mu
}\right) \delta \vec{n}_{i}\delta \vec{n}_{j}
\label{E2}
\end{equation}%
and%
\begin{equation}
\delta E_{ij}^\mathrm{AT}=-4\sum_{\mu ,\nu =x,y,z}\delta n_{i}^{\mu }A_{ij}^{\mu \nu
}\delta n_{j}^{\nu }\;,  \label{E3}
\end{equation}%
i.e., we obtain a Heisenberg-type (HT) and an anisotropic-type (AT) term expressed by generalized exchange parameters, as given in Eq. (\ref{Adeff}). The details of the derivation of Eqs. (\ref{E1})-(\ref{E3}) can be found in Appendix B.

As we mentioned earlier, only parameters $A_{ij}^{00 }$ and $A_{ij}^{zz }$ should be considered in the collinear limit, therefore for this case, the two-site energy variation formula simplifies to
\begin{equation}
\delta E_{ij}^\mathrm{two}=-2J_{ij}^\mathrm{L} \left(\delta 
{n}_{i}^{x}\delta {n}_{j}^{x} +\delta 
{n}_{i}^{y}\delta {n}_{j}^{y}\right)-2
G_{ij} \delta {n}_{i}^{z
}\delta {n}_{j}^{z}\;, \label{finall}
\end{equation}%
where 
$G_{ij}=A_{ij}^{00}+A_{ij}^{zz}$, implying that we have to deal with two parameters 
to describe the exchange interaction even in case of collinear spin arrangement. The parameter $J^\mathrm{L}_{ij}$ describes the transversal ($x$ or $y$) part of the energy variation and the longitudinal ($z$) part is characterized by the parameter $G_{ij}$, which are proportional to $\left(\delta \theta \right)^{2}$ and $\left(\delta \theta \right)^{4}$, respectively. 
In the collinear limit it is sufficient to keep only the HT term in Eq. (\ref{E1}) as was done in Ref.~\cite{oldlicht}. In this case $\delta E_{ij}^\mathrm{two}$ equals $\delta E_{ij}^\mathrm{HT}$ which can be briefly written as $-2J_{ij}^\mathrm{L}\,\delta \vec{n}_{i}\delta \vec{n}_{j}$. {By limiting to the {\em bilinear  scalar} Heisenberg effective spin model with exchange parameter $J_{ij}^\mathrm{L}$ we also recover in the collinear limit the energy variation described by Eq. (\ref{onelicht}), as derived in Appendix C, a result that is in agreement with Ref. \cite{oldlicht}. 

However, in order to keep both HT and AT terms in the general two-site MSF energy deviation formula (\ref{E1}), we attempt to map the MSF parameters onto a {\em bilinear tensorial} effective Hamiltonian
\begin{equation}
\mathcal{H}^\mathbf{T}=-\sum_{ij}^{i\neq j}\vec{n}_{i}%
\mathbf{J}_{ij}\vec{n}_{j}\;, \label{Htens}
\end{equation}%
with a tensor interaction $\mathbf{J}_{ij}=\{J_{ij}^{\mu \nu };\mu,\nu\in\{x,y,z\}\}$. 
It can easily be seen that the two-site energy variation can be written as $-2\delta \vec{n}%
_{i}\mathbf{J}_{ij}\delta \vec{n}_{j}$. We compare this expression with Eqs. (\ref{E1})-(\ref{E3}) and identify that 
\begin{equation}
J_{ij}^{\mu \nu }=\left(
A_{ij}^{00}-A_{ij}^{xx}-A_{ij}^{yy}-A_{ij}^{zz}\right) \delta _{\mu \nu
}+2A_{ij}^{\mu \nu }\;,  \label{gennn}
\end{equation}%
which recovers Eq. (\ref{finall}) for collinear systems. These results are similar but not identical to those of Ref. \cite{antro1} and \cite{antro2}, where e.g.~the corresponding expression  to Eq. (\ref{gennn}) was written as $J_{ij}^{\mu \nu }=A_{ij}^{00} \delta _{\mu \nu}-A_{ij}^{\mu \nu }$.
If we now use Eq. (13) to obtain the energy for the collinear one-site rotation, we obtain $-2\sum\limits G_{ji} \delta {n}_{i}^{z} $, which is not consistent with the expression of Eq. (\ref{onelicht}), see Appendix C for a derivation. The fact that Eq. (\ref{onelicht}) is not recovered in this case, implies that one {\em cannot} map the non-collinear MSF parameters onto a tensorial effective Hamiltonian, as formulated in Eqs. (\ref{Htens})-(\ref{gennn}). 

This motivates to take an alternate approach and consider higher order spin terms in the spin Hamiltonian, in the spirit of Ref.~\cite{lounis}. The simplest extension is the {\em biquadratic} effective Hamiltonian 
\begin{equation}
\mathcal{H}^\mathrm{Q}=-\sum_{ij}^{i\neq j}J_{ij}^{\prime }%
\vec{n}_{i}\vec{n}_{j}-\sum_{ij}^{i\neq j}B_{ij}\left( 
\vec{n}_{i}\vec{n}_{j}\right) ^{2}\;,  \label{biquad}
\end{equation}
where a revised bilinear parameter $J_{ij}^{\prime}$ is introduced besides the biquadratic one $B_{ij}$.  
Deriving the two-site energy variation formula from Eq.~(\ref{biquad}) we obtain that the biquadratic two-site rotation energy deviation can be written as a sum of a biquadratic Heisenberg-type term (QHT) and anisotropic-type term (QAT), where
 \begin{equation}
\delta E_{ij}^\mathrm{QHT}=-2\left[ J_{ij}^{\prime }+2B_{ij} \left( \vec{n}_{i} \vec{n}_{j} \right) \right] \delta \vec{n}_{i}\delta \vec{n}_{j}\; \label{HT}
\end{equation}%
and
 \begin{equation}
\delta E_{ij}^\mathrm{QAT}=-4B_{ij} \sum_{\mu ,\nu =x,y,z}\delta n_{i}^{\mu } n_{i}^{\nu} n_{j}^{\mu} \delta n_{j}^{\nu }\;, \label{AT}
\end{equation}%
when we consider the $\left(\delta n_{i(j)}^{\mu}\right)^{2}$-type terms in the two-site rotation energy variation formulas, see Appendix C for a derivation. 
It should be noted that 
Eq.~(\ref{HT}), reduces to $-\left(2J_{ij}^{\prime }+4B_{ij} \right) \delta \vec{n}_{i}\delta \vec{n}_{j}$ while
Eq.~(\ref{AT}), can be written as $-4B_{ij} \delta n_{i}^{z} \delta n_{j}^{z}$ in case of the (ferromagnetic) collinear limit. Comparing Eqs.~(\ref{E2}) and (\ref{E3}) with Eqs. (\ref{HT}) and (\ref{AT}) 
one can identify 
\begin{equation}
J_{ij}^{\prime }=A_{ij}^{00}-3A_{ij}^{zz} \;,\;\;\;\:\;\;\;B_{ij}=A_{ij}^{zz} \;. \label{asmp1}
\end{equation}%
In case of one-site rotations the leading term can be written as $\delta E_{i}^\mathrm{one}-2\sum \left(J_{ji}^{\prime }+2B_{ji} \right)\delta n_{i}^{z}$ where $J_{ji}^{\prime} + 2B_{ji}=A_{ji}^{00}-A_{ji}^{zz}=J_{ji}^\mathrm{L}$, i.e. Eq.~(\ref{onelicht}) has been recovered. This is a required condition for any effective spin-Hamiltonian, since the analysis from multiple scattering theory establishes Eq. (\ref{onelicht}). Hence, the recovery of Eq. (\ref{onelicht}) when considering one-site rotations, shows that the collinear MSF parameters {\em can} be mapped onto a biquadratic model.

\begin{figure}[h!]
\begin{center}
\begin{tabular}
[c]{c}%
\includegraphics[width=0.8\textwidth, bb= 0 10 1250 510]{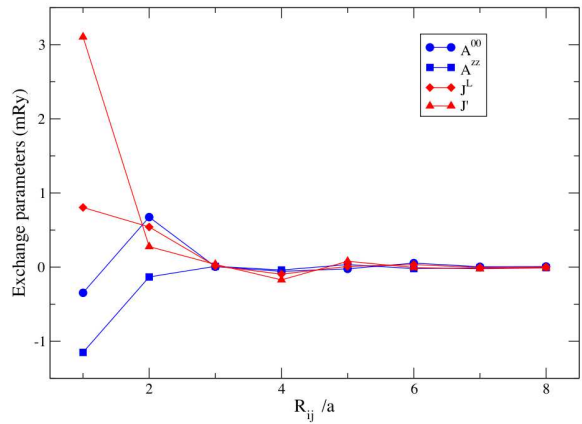}
\end{tabular}
\end{center}
\par
\vskip -0.5cm  \caption{Collinear exchange parameters ($A^{00}$ and $A^{zz}$) between the first eight neighbors in bcc Fe. $J^\mathrm{L}$, see Eq. (\ref{lichtdef}) and $J'$, see Eq. (\ref{asmp1}) are derived parameters for the bilinear and biquadratic spin Hamiltonians, the biquadratic $B$ equals $A^{zz}$, which is rather large for nearest neighbors.}%
\label{Fefig}%
\end{figure}

\begin{figure}[h!]
\begin{center}
\begin{tabular}
[c]{c}%
\includegraphics[width=0.8\textwidth, bb= 0 10 1250 510]{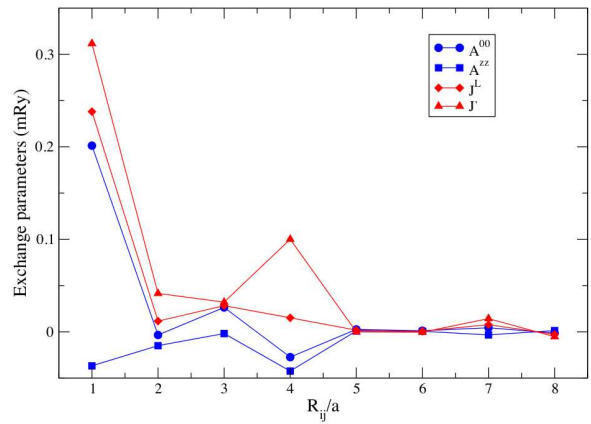}
\end{tabular}
\end{center}
\par
\vskip -0.5cm  \caption{Collinear exchange parameters between the first eight neighbors in fcc Ni. The bilinear parameter of bilinear model, $J^\mathrm{L}$, and the bilinear parameter of biquadratic model, $J'$, are very close to each other in fcc Ni, $A^{zz}\ll A^{00}$.}%
\label{Nifig}%
\end{figure}

The numerical calculations of these parameters have been implemented in terms of a real-space (RS-)LMTO-ASA code, see Ref. \cite{anders}. The LMTO formalism used in this work and its connection to MSF is discussed in Ref. \cite{andersen2}, and it has been shown that it results in LKAG parameters which are consistent with other electronic structure methods \cite{frota}. The calculated exchange parameters between the first nearest neighbor sites are shown in Fig. \ref{Fefig} and \ref{Nifig} for bcc Fe and fcc Ni, respectively. We obtained that $A^{zz}_{ij}=B_{ij}$ is much larger than $A^{00}_{ij}$  for nearest neighbors in bcc Fe . Also, $J^\mathrm{L}_{ij}$ is drastically different than $J_{ij}^{\prime}$ showing the importance of higher order spin interactions, see Fig. \ref{Fefig}. It might seem, from Fig. \ref{Fefig}, that the biquadratic Hamiltonian (with  $J_{ij}^{\prime}$ and $B_{ij}$)  and the bilinear Hamiltonian (with $J_{ij}^{L}$) give different excitation energies. In the collinear limit they actually give rise to the same excitation energies, since, as follows from Eqs. (\ref{lichtdef}) and (\ref{asmp1}), the relationships $J^\prime_{ij} + 2 B_{ij} = J^L_{ij}$ and, consequently, $\delta E_{ij}^{\mathrm{HT}} = \delta E_{ij}^{\mathrm{QHT}}$ hold for all pairs. On the other hand, in case of fcc Ni first neighbor pairs, the $A_{ij}^{zz}$ and, therefore, the biquadratic parameter are very small, so that $J^\mathrm{L}_{ij}$ and $J_{ij}^{\prime}$ are close to each other as shown in Fig. \ref{Nifig}. Fig. \ref{Nifig} shows that $A_{ij}^{zz}$ (hence $B_{ij}$) deviates from the general trend, in the case of fourth nearest neighbor interaction. This is counterbalanced by a larger value of $J_{ij}^{\prime}$ for this interaction-distance. Hence, also in this case will the biquadratic Hamiltonian and the bilinear Hamiltonian give rise to the same excitation energies, in the collinear limit. The data in Fig. \ref{Nifig} results in a lower value of $B_{ij}$, which implies that the bilinear term is more dominating for fcc Ni. It should be noted that the bilinear and the biquadratic model result in the same magnon spectra in case of long wavelengths, but for a more accurate mapping procedure one should consider third and fourth order type terms of $\delta n_{i(j)}^{\mu}$ in the two-site rotation energy variation formula. We also note that Eq. (\ref{finall}), which is a special (collinear) case of the general non-collinear expression in Eqs. (\ref{E1}-\ref{E3}), was discussed recently in Ref.~\cite{lounis}, and in order to map it into a spin model
 a four-spin model was introduced.


\begin{figure}[h!]
\begin{center}
\begin{tabular}
[c]{c}%
\includegraphics[width=0.8\textwidth, bb= 0 10 1200 510]{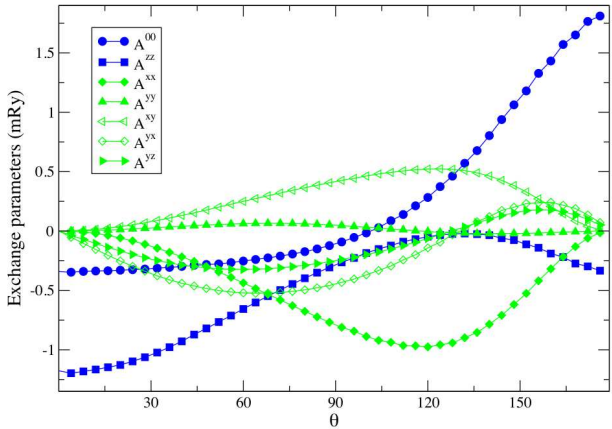}
\end{tabular}
\end{center}
\par
\vskip -0.5cm  \caption{Comparing the collinear exchange parameters (blue circles and blue squares) and non-collinear parameters (green diamonds and triangles) two nearest neighbor sites of bcc Fe as a function of angle rotating one magnetic moment with angle $\theta$.}%
\label{ROT}%
\end{figure}

\begin{figure}[h!]
\begin{center}
\begin{tabular}
[c]{c}%
\includegraphics[width=0.8\textwidth, bb= 0 10 1200 510]{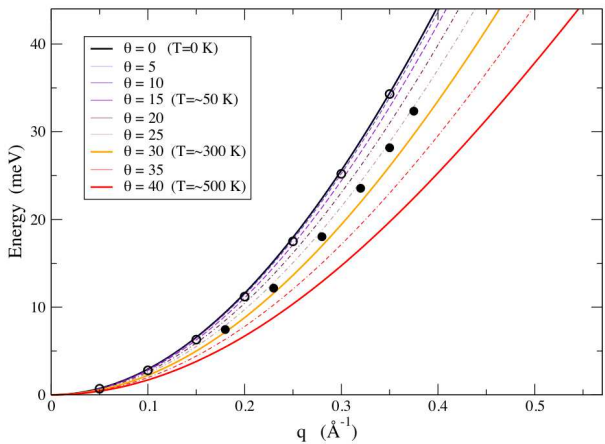}
\end{tabular}
\end{center}
\par
\vskip -0.5cm  \caption{Spin-wave dispersion relation calculated at different temperatures along the $\Gamma$-$H$ direction. The top thick (black) line shows the calculated spectrum from collinear LKAG exchange parameters, open circles come from magnetization measurement at 4.2 K \cite{exp1} The middle thick (yellow) line corresponds to the calculated spectrum at 300 K and the filled circles refer to the room temperature neutron scattering measurement data\cite{Lynn}. }%
\label{OFFROT}%
\end{figure}

Most importantly, our formulation allows to consider non-collinear spin configurations and to calculate magnon spectra from these configurations. In Fig. \ref{ROT} we show the calculated exchange parameters between two nearest neighbors, when rotating only one spin of a bcc Fe lattice. It can be seen that the collinear parameters are decreasing and the non-collinear parameters are increasing when $\theta$ increases. We note that the configuration dependence of $A^{zz}$ is much stronger compared to $A^{00}$ as shown in Fig. \ref{ROT}. As we mentioned earlier, the $A^{zz}$ is the dominant term in case of first neighbor pairs while $A^{00}$ gives the main contribution of exchange couplings for more long ranged neighbors, see Fig. \ref{Fefig}, therefore the configuration dependence of the second and further neighbor LKAG parameters is negligible.

In order to estimate how the finite temperature induced spin-order of the lattice influences the exchange interaction, and the magnetic excitation energies, we performed Monte Carlo simulations (with 128 Fe atoms) using the parameters of Eq. (\ref{lichtdef}), and we performed a statistical analysis of the distribution of angles between the spins on the simulation box. We find that at 300 K the average deviation of an atomic spin-moment from the global quantization axis is $\hat{\theta}=28\,^{\circ}$, while in case of 500 K this deviation was $\hat{\theta}=39\,^{\circ}$. We then performed a calculation of the parameters in Eqs. (\ref{Adeff}), (\ref{asmp1}) using a spin-configuration with deviations ($\hat{\theta}$-s) from the global magnetization direction, with angles given by the Monte Carlo simulations (e.g. $28\,^{\circ}$ at room temperature). Similarly to the case when only one spin-moment was rotated, the non-collinear parameters become significant for larger average spin-moment deviation. We then performed a statistical analysis of these parameters and obtained averages over different site- and $\mu$, $\nu$-indices. We analyzed these average exchange parameters in case of $\hat{\theta}=0\,^{\circ}$,  $5\,^{\circ}$,  $10\,^{\circ}$, $25\,^{\circ}$ and $40\,^{\circ}$. For small angles, the collinear parameters are, as expected, dominant. At room temperature the non-collinear parameters are roughly 30 percent of the collinear ones, and  at 500 K, i.e. in case of $40\,^{\circ}$ average angles between atomic and global magnetization direction, the collinear and non-collinear parameters are of the same order. We then calculated the spin wave spectra along the $\Gamma$-$H$ direction, for finite temperature configurations, the result is shown in Fig. \ref{OFFROT}. These spin-wave spectra are obtained from configuration dependent LKAG exchange parameters evaluated for non-collinear configurations corresponding to temperatures ranging from 0 to 500 K. As Fig. \ref{Fefig} shows that the first and second neighbor values are dominant, we evaluated the spectra from these parameters only. At zero temperature ($\theta=0$) we obtained a spin-wave stiffness constant of 287 meV\AA$^{2}$, whereas a magnetization measurement at 4.2 K resulted in $D_\mathrm{exp}=$ 280-330 meV\AA$^{2}$ \cite{exp1}, \cite{exp2}. In Fig. \ref{OFFROT} we compare our theory with the experimental data of Ref. \cite{exp1}, where the top thick (black) line shows the calculated spectrum from collinear LKAG exchange parameters, and the open circles represent experimental data. These experimental values were evaluated from the experimental spin-wave stiffness constant, using the expression $D_\mathrm{exp}q^{2}$, and it may be seen that experiment and theory agree. Ref. \cite{exp3} has carefully examined the temperature dependence of magnetic excitations of iron from neutron scattering data, and the measured room temperature spin-wave spectrum of bcc Fe is shown in Fig. \ref{OFFROT} by filled black circles. It is found that experimental values are close to our calculated room temperature (yellow) curve. The experimental room temperate spin stiffness value is 230, to be compared to our calculated value of 219 meV\AA$^{2}$. Furthermore, Fig. 2. in Ref. \cite{exp3} shows measured spectra along the (110) direction, starting from low values and increasing T up to the Curie temperature. The measured trend is obvious, softer curves are observed with increasing temperature, in a fashion which is similar to our calculations (Fig. \ref{OFFROT}).

In conclusion, we have derived a general exchange coupling expression, which treats also non-collinear spin-orientations, to describe the interaction between magnetic moments. This formula contains an anisotropic-type term even in the absence of spin-orbit coupling, which leads to different transversal and longitudinal exchange parameters, that survive even in collinear systems. Keeping this term we have demonstrated that one should map these parameters onto a higher order (biquadratic) spin Hamiltonian, which results in the same effective field as the LKAG scalar Heisenberg model. From numerical calculations we have shown that this anisotropic parameter, i.e., the biquadratic coupling, is rather large in bcc Fe and quite small in fcc Ni. Nevertheless, this term is in the collinear case less important since it has energy contributions $\delta \theta ^{4}$. We have shown that in non-collinear spin-configurations this anisotropic term is described by a tensor and we have examined the configuration dependence of its elements going from a collinear to non-collinear states. We have obtained, on one hand, that effects of non-collinearity on these parameters are quite small at low temperature but that for finite temperatures the influence of non-collinearity on the exchange interactions becomes significant. We have calculated the bcc Fe spin stiffness constant and the magnon spectra at different temperatures quantitatively recovering the finite temperature experimental data. Our work opens  up for a truly ab-initio description of finite temperature effects of the interatomic exchange, end hence enables accurate finite temperature, spin-dynamics simulations of magnetic materials.

The authors owe thanks to Corina Etz, Adam Jakobsson, L\'aszl\'o Udvardi and Patrik Thunstr\"om for the fruitful discussions. This work has been financed by eSSENCE, the KAW foundation, VR and ERC (project 247062 - ASD). Financial support was in part provided by the New Sz\'echenyi Plan of Hungary (Project ID.~T\'AMOP-4.2.2.B-10/1--2010-0009) and the Hungarian Scientific Research Fund (contract OTKA 83114, OTKA K77771 and IN83114).

\end{document}


\title{Interatomic exchange interactions in non-collinear magnets}
\author{A. Szilva}
\affiliation{Department of Physics and Astronomy, Division of Materials Theory, Uppsala University,
Box 516, SE-75120, Uppsala, Sweden}
\author{M. Costa}
\affiliation{Department of Physics and Astronomy, Division of Materials Theory, Uppsala University,
Box 516, SE-75120, Uppsala, Sweden}
\affiliation{Instituto de F\'isica, Universidade Federal Fluminense, 24210-346 Niter\'oi, Rio de Janeiro, Brazil}
\affiliation{Department of Physics and Astronomy, University of California, Irvine, California 92697, USA}
\author{A. Bergman}
\affiliation{Department of Physics and Astronomy, Division of Materials Theory, Uppsala University,
Box 516, SE-75120, Uppsala, Sweden}
\author{L. Szunyogh}
\affiliation{Department of Theoretical Physics and Condensed Matter Research Group of Hungarian Academy of Sciences, Budapest University of Technology and Economics, Budafoki \'ut 8. H1111 Budapest, Hungary}
\author{L. Nordstr\"om}
\affiliation{Department of Physics and Astronomy, Division of Materials Theory, Uppsala University,
Box 516, SE-75120, Uppsala, Sweden}
\author{O. Eriksson}
\affiliation{Department of Physics and Astronomy, Division of Materials Theory, Uppsala University,
Box 516, SE-75120, Uppsala, Sweden}

\begin{abstract}

\end{abstract}
\pacs{later}

\maketitle

\section{Appendix A - one-site rotations}

In the spirit of the magnetic force theorem \cite{meth}, the grand canonical potential
\begin{equation}
E=-\int\limits_{-\infty }^{\infty}d\varepsilon 
N(\varepsilon ) f({\varepsilon})\;  \label{grandcan}
\end{equation}%
is used to characterize the energy of the many-electron system, where $N({\varepsilon})$ denotes the integrated density of states (integrated DOS) and $f({\varepsilon})$ stands for the Fermi-Dirac distribution. Approaching $f(\varepsilon)$ by a step function well below the Fermi temperature, the variation of the energy can be written as 
\begin{equation}
\delta E=-\int\limits_{-\infty }^{\varepsilon _{F}}d\varepsilon \delta
N(\varepsilon )\;  \label{deltaen}
\end{equation}%
in case of any perturbation. The so-called Lloyd formula expresses how the integrated DOS can be calculated in the presence of any perturbation. If $N_{0}(\varepsilon )$ denotes the integrated DOS calculated from the (potential) free Hamiltonian then the integrated DOS can be written as
\begin{equation}
N(\varepsilon )=N_{0}(\varepsilon )+\frac{1}{\pi }\operatorname{Im}%
Tr_{\sigma L}\ln \tau \left( \varepsilon \right) \;,
\end{equation}%
where $N(\varepsilon )$-$N_{0}(\varepsilon )$ basically comes from the sum of the scattering phase shifts, i.e., it is the famous Friedel sum-rule. In case of one-site perturbation, when a spin moment at site $i$ is rotated by infinitesimal angle $\delta \theta$, the scattering path
operator can be denoted by $\tau _{ii}^{\prime }$, where
\begin{equation}
\tau _{ii}^{\prime }=\tau _{ii}\left( I_{2}+\delta
P_{i}\tau _{ii}\right) ^{-1}\;,
\end{equation}
where $\delta P_{i}=P_{i}^{\prime }-P_{i}$ and $P_{i}^{\prime }$
stands for the perturbed inverse single site scattering matrix. Applying the Lloyd formula for infinitesimal one-site rotation (perturbation), the integrated DOS can be obtained as%
\begin{equation}
N^{\prime}(\varepsilon )=N_{0}(\varepsilon )+\frac{1}{\pi }\operatorname{Im}Tr_{\sigma L}\ln
\tau \left( \varepsilon \right) +\delta N^{one}_{i}\;,  \label{N}
\end{equation}%
where
\begin{equation}
\delta N^{one}_{i} =-\frac{1}{\pi }\operatorname{Im}Tr_{\sigma L}\ln \left( I_{2}+\delta
P_{i}\tau _{ii}\right) \;. \label{oneN}
\end{equation}
After substituting into Eq. (\ref{deltaen}) we obtain%
\begin{equation}
\delta E_{i}^{one}=\frac{1}{\pi }\int\limits_{-\infty }^{\varepsilon
_{F}}d\varepsilon \operatorname{Im}Tr_{\sigma L}\ln \left( I_{2}+\delta P_{i}\tau_{ii} \right) \;,
\label{eq}
\end{equation}
where the expression $Tr_{\sigma L}\ln \left( I_{2}+\delta P_{i}\tau_{ii} \right)$ can be explicitly written as
\begin{equation}
Tr_{\sigma L}\ln \left( 1+p_{i}\left( \delta \vec{n}_{i}\vec{T}_{ii}\right) I_{2}+\left(
p_{i}T_{ii}^{0}\delta \vec{n}_{i}+ip_{i}\left( \delta \vec{n}_{i}\times \vec{%
T}_{ii}\right) \right) \vec{\sigma} \right) \; \label{blabla}
\end{equation}
and we will use the brief notation $X$ for $\delta P_{i}\tau_{ii}$. Below we discuss this expression for the collinear case and the general non-collinear case. 

\subsection{Collinear scheme ($T_{ij}^{x}=T_{ij}^{y} \simeq 0$)}

In this case expression Eq. (\ref{blabla}) simplifies as follows,
\begin{equation}
Tr_{\sigma L}\ln \left( 1+p_{i} {T}_{ii}^{z}\delta n_{i}^{z}+
p_{i}T_{ii}^{0}\delta \vec{n}_{i} \vec{\sigma}+ip_{i}{T}_{ii}^{z}\delta n_{i}^{y} \sigma^{x}-ip_{i}{T}_{ii}^{z}\delta n_{i}^{x} \sigma^{y} \right)   \;
\end{equation}
i.e., we can write
\begin{equation}
I_{2}+X =\left(
\begin{array}
[c]{cccc}%
1+p_{i} {T}_{ii}^{\uparrow}\delta n_{i}^{z}  & p_{i} {T}_{ii}^{\downarrow}\delta n_{i}^{x}-ip_{i} {T}_{ii}^{\downarrow}\delta n_{i}^{y}  \\
p_{i} {T}_{ii}^{\uparrow}\delta n_{i}^{x} +ip_{i} {T}_{ii}^{\uparrow}\delta n_{i}^{y}& 1-p_{i} {T}_{ii}^{\downarrow}\delta n_{i}^{z}   \\
\end{array}
\right) \; \label{coll}
\end{equation}
using definitions $T_{ii}^{\uparrow}=T_{ii}^{00}+T_{ii}^{zz}$ and $T_{ii}^{\downarrow}=T_{ii}^{00}-T_{ii}^{zz}$. Considering the relation
\begin{equation}
Tr_{\sigma L} \ln{\left(I_{2}+X \right)}=Tr_{L} \ln \det{\left(I_{2}+X \right)} \;,
\end{equation}
it can be obtained that
\begin{equation}
Tr_{\sigma L}\ln \left(I_{2}+X \right)=Tr_{ L}\ln \left(1+x  \right)   \;,
\end{equation}
where
\begin{equation}
x=p_{i} \left({T}_{ii}^{\uparrow}-{T}_{ii}^{\downarrow}\right) \delta n_{i}^{z}-
p_{i}T_{ii}^{\uparrow}p_{i}T_{ii}^{\downarrow}  \delta \vec{n}_{i} \delta \vec {n}_{i}   \;
\end{equation}
or
\begin{equation}
x=2p_{i} T_{ii}^{z} \delta n_{i}^{z}-
p_{i}\left({T}_{ii}^{0}+{T}_{ii}^{z}\right)p_{i}\left({T}_{ii}^{0}-{T}_{ii}^{z}\right)  \delta \vec{n}_{i} \delta \vec {n}_{i}   \;.
\end{equation}
Here, we use that $\delta n_{i}^{x}=\sin {\delta \theta} \simeq \delta \theta$ and assume that $\delta n_{i}^{y}=0$, i.e., we choose the coordinate system such that the spin moment is rotated in the $xz$-plane as in Ref. \cite{oldlicht}. Since $\delta n_{i}^{z}=\cos {\delta \theta}-1 \simeq \left( -1/2  \right)\left( \delta \theta \right)^{2}$, we can write that $ \delta \vec{n}_{i} \delta \vec {n}_{i} =\delta n_{i}^{z}\delta n_{i}^{z}+\delta n_{i}^{x}\delta n_{i}^{x}+\delta n_{i}^{y}\delta n_{i}^{y} \simeq-2\delta n_{i}^{z}$. Then we obtain
\begin{equation}
Tr_{\sigma L}\ln \left(I_{2}+X \right) =Tr_{ L}\ln \left(1+x \right) \simeq Tr_{ L} \left(x \right)  \;,
\end{equation}
where
\begin{equation}
x= 2 \left( p_{i}   \frac{{T}_{ii}^{\uparrow}-{T}_{ii}^{\downarrow}}{2 } +
p_{i}T_{ii}^{\uparrow}p_{i}T_{ii}^{\downarrow}  \right)\delta n_{i}^{z}  \;.
\end{equation}
Substituting this result into Eq. (\ref{eq}) we can write that
\begin{equation}
\delta E^{one}_{i}\simeq \frac{2}{\pi }\int\limits_{-\infty }^{\varepsilon
_{F}}d\varepsilon \operatorname{Im}Tr_{L}\left( p_{i}
T_{ii}^{z} + p_{i}T_{ii}^{\uparrow}p_{i}T_{ii}^{\downarrow} \right) \delta {n}_{i}^{z} \;.  \label{one3}
\end{equation}%
Considering Eqs. (1), (2) and (3) in the main paper we obtain that
\begin{equation}
\frac{1}{2}\left( T_{\uparrow }^{-1}-T_{\downarrow }^{-1}\right)
_{ij}=p_{i}\delta _{ij} \;,
\end{equation}%
which implies 
\begin{equation}
\frac{T_{ii}^{\uparrow }-T_{ii}^{\downarrow }}{2}=-\sum\limits_{j}T_{ij}^{%
\uparrow }p_{j}T_{ji}^{\downarrow }\;,  \label{sum}
\end{equation}%
i.e., we have derived a sum rule. Substituting this sum rule into Eq. (\ref{one3}), the energy deviation of the one-site rotation can be obtained as
\begin{equation}
\delta E^{one}_{i}=-2\sum\limits_{j (\neq i)} \left(A^{00}_{ij}-A^{zz}_{ij}\right) \delta {n}_{i}^{z} =-2\sum\limits_{j (\neq i)}J_{ij}^{L},  \; \label{onelicht}
\end{equation}%
i.e., we obtain the same result as that of Ref. \cite{oldlicht} and Eq. (7) of our paper.
 
\subsection{General scheme ($T_{ij}^{x}=T_{ij}^{y} \neq 0$)} 
  
We would now like to determine the energy deviation of the one-site spin rotation in non-collinear configurations. The collinear scheme can be applied when the spin deviation is close to the the collinear ground state. Otherwise the terms $T^{x}_{ij}$ or $T^{y}_{ij}$ can have important contributions in Eqs. (\ref{eq}) and (\ref{blabla}). We can write again that
\begin{equation}
Tr_{\sigma L} \ln{\left(I_{2}+X \right)}=Tr_{L} \ln \det{\left(I_{2}+X \right)} \;,
\end{equation}
i.e.,
\begin{equation}
Tr_{\sigma L}\ln \left(I_{2}+X \right)=Tr_{ L}\ln \left(1+x  \right)   \;, \label{xdef}
\end{equation}
but now
\begin{equation}
x=2p_{i} \vec{T}_{ii} \delta \vec {n}_{i}-a_{i}^{0,0} \delta \vec{n}_{i}\delta \vec{n}_{i}+\sum\limits_{\mu=x,y,z}a_{i}^{\mu,\mu}\delta n_{i}^{\mu}\delta n_{i}^{\mu}+   \; \label{xresult}
\end{equation}
\begin{equation}
+2\sum\limits_{\mu, \nu (\mu \neq \nu)} a_{i}^{\mu,\mu} \delta n_{i}^{\nu}\delta n_{i}^{\nu}-2\sum\limits_{\mu, \nu (\mu \neq \nu)} a_{i}^{\mu,\nu} \delta n_{i}^{\mu}\delta n_{i}^{\nu}   \;, \label{xresult2}
\end{equation}
where
\begin{equation}
a_{i}^{\alpha, \beta}=p_{i} {T}_{ii}^{\alpha}p_{i} {T}_{ii}^{\beta}   \;. \label{adeff}
\end{equation}
It is in principle possible to calculate the one-site energy deviation from Eqs. (\ref{xdef})-(\ref{adeff}), to obtain a formula for the non-collinear effective (Weiss) field.

\section{Appendix B - two-site rotations}
 
We rotate two spin moments at site $i$ and $j$ oppositely with angle $\delta \theta$ at the same time. The perturbed SPO can be calculated as
\begin{equation}
\tau _{ij}^{\prime }=\tau _{ij}\left( I_{2}+\delta P_{i}\tau _{ij}+\delta
P_{j}\tau _{ji}\right) ^{-1}\;,
\end{equation}%
which implies that the total variation of the integrated DOS can be written as $\delta N^{one}_{i}+\delta N^{one}_{j}+\delta N_{ij}^{two}$, where $\delta N^{one}_{i(j)}$ is defined by Eq. (\ref{oneN}), and
\begin{equation}
\delta N_{ij}^{two} =-\frac{1}{\pi }\operatorname{Im}Tr_{\sigma L}\ln \left\{
Y_{ij}\right\} \;,  \label{dNtwo}
\end{equation}
where
\begin{equation}
Y_{ij}=I_{2}-\left( I_{2}+\delta P_{i}\tau _{ii}\right) ^{-1}\delta P_{i}\tau _{ij}\delta
P_{j}\tau _{ji}\left( I_{2}+\delta P_{j}\tau _{jj}\right) ^{-1}\;.
\end{equation}%
If the total energy of the perturbed (or rotated) system is denoted by $E^{\prime}$, and $E$ stands for the energy of the unperturbed system then
\begin{equation}
E^{\prime}=E+\delta E^{one}_{i}+\delta E^{one}_{j}+\delta E^{two}_{ij}
\end{equation}
in case of two-site rotation, where
\begin{equation}
\delta E^{two}_{ij}=-\int\limits_{-\infty }^{\varepsilon _{F}}d\varepsilon \delta
N^{two}_{ij}(\varepsilon )\;.  \label{deltaentwo}
\end{equation}%
A new term $\delta E^{two}_{ij}$ has appeared in the $E^{\prime}$-$E$ energy variation expression, which describes the interaction between magnetic spin moment at site $i$ and $j$. We will analyze the expression of the interaction term in this Section.

Using $\ln (1+x)\simeq x$ we find%
\begin{equation}
\delta N_{ij}^{two}\simeq \frac{1}{\pi }\operatorname{Im}Tr_{\sigma L}\left( \delta
P_{i}\tau _{ij}\delta P_{j}\tau _{ji}\right)
\end{equation}%
in leading order, i.e., we obtain%
\begin{equation}
\delta E_{ij}^{two}\simeq -\frac{1}{\pi }\int\limits_{-\infty
}^{\varepsilon _{F}}d\varepsilon \operatorname{Im}Tr_{\sigma L}\left( \delta
P_{i}\tau _{ij}\delta P_{j}\tau _{ji}\right) \;,  \label{deltae}
\end{equation}%
after substituting expression (\ref{dNtwo}) into Eq. (\ref{deltaentwo}). We evaluate the trace expression in the obtained equation using the relations
\begin{eqnarray}
\vec{a}\left( \vec{b}\times \vec{c}\right) &=&\vec{b}\left( \vec{c}\times 
\vec{a}\right) =\vec{c}\left( \vec{a}\times \vec{b}\right) \\
\left( \vec{a}\times \vec{b}\right) \left( \vec{c}\times \vec{d}\right)
&=&\left( \vec{a}\vec{c}\right) \left( \vec{b}\vec{d}\right) -\left( \vec{d}%
\vec{a}\right) \left( \vec{b}\vec{c}\right) \\
Tr(ABCD) &=&Tr(BCDA)=Tr(CDAB) \;.
\end{eqnarray}%
We should evaluate $Tr_{\sigma L}$
\begin{equation*} 
\begin{array}{c}
\left( p_{i}\vec{T}_{ij}\delta \vec{n}_{i}\right) \left( p_{j}\vec{T}%
_{ji}\delta \vec{n}_{j}\right) I_{2}+ \\ 
\left( p_{i}T_{ij}^{0}\delta \vec{n}_{i}+ip_{i}\delta \vec{n}_{i}\times \vec{%
T}_{ij}\right) \left( p_{j}T_{ji}^{0}\delta \vec{n}_{j}+ip_{j}\delta \vec{n}%
_{j}\times \vec{T}_{ji}\right) I_{2}+ \\ 
\left( p_{i}T_{ij}^{0}\delta \vec{n}_{i}+ip_{i}\delta \vec{n}_{i}\times \vec{%
T}_{ij}\right) \left( p_{j}\vec{T}_{ji}\delta \vec{n}_{j}\right) \vec{\sigma}%
+ \\ 
\left( p_{i}\vec{T}_{ij}\delta \vec{n}_{i}\right) \left(
p_{j}T_{ji}^{0}\delta \vec{n}_{i}+ip_{j}\delta \vec{n}_{j}\times \vec{T}%
_{ji}\right) \vec{\sigma}+ \\ 
i\left( p_{i}T_{ij}^{0}\delta \vec{n}_{i}+ip_{i}\delta \vec{n}_{i}\times 
\vec{T}_{ij}\right) \times \left( p_{j}T_{ji}^{0}\delta \vec{n}%
_{j}+ip_{j}\delta \vec{n}_{j}\times \vec{T}_{ji}\right) \vec{\sigma}%
\end{array}%
\end{equation*}%
which can be written as%
\begin{equation}
2Tr_{L}\left[ 
\begin{array}{c}
\left( p_{i}T_{ij}^{0}p_{j}T_{ji}^{0}-p_{i}\vec{T}_{ij}p_{j}\vec{T}%
_{ji}\right) \left( \delta \vec{n}_{i}\delta \vec{n}_{j}\right) + \\ 
\left( p_{i}T_{ij}^{\mu }p_{j}T_{ji}^{\nu }+p_{j}T_{ji}^{\nu
}p_{i}T_{ij}^{\mu }\right) \delta n_{i}^{\mu }\delta n_{j}^{\nu } \\ 
i\left( p_{i}T_{ij}^{0}p_{j}\vec{T}_{ji}-p_{j}T_{ji}^{0}p_{i}\vec{T}%
_{ij}\right) \left( \delta \vec{n}_{i}\times \delta \vec{n}_{j}\right)%
\end{array}%
\right] \;,
\end{equation}%
where $\mu $ and $\nu $ run over $x$, $y$ and $z$. Using Eqs. (1)-(3) in our paper, we can prove that $T^{\alpha}_{ij}=T^{\alpha}_{ji}$, $\alpha \in\{0,x,y,z\}$, in the absence of spin-orbit coupling, which implies that $\mathrm{Tr}\left( p_{i}T_{ij}^{0}p_{j}\overrightarrow{T}_{ji}\right)=\mathrm{Tr}\left( p_{j}T_{ji}^{0}p_{i}%
\overrightarrow{T}_{ij}\right)$, therefore the  $\delta \vec{n}_{i}\times \delta \vec{n}_{j}$ (Dzyaloshinskii-Moriya-type) term does not give a contribution. We obtain the final result for energy deviation in case of the two-site spin rotations, and it is written in Eqs. (8), (9) and (10) in the paper.

\section{Appendix C - Effective spin Hamiltonians}

\subsection{Bilinear (scalar) Heisenberg model}

For this model we write 
\begin{equation}
\mathcal{H}^{L} =-\sum_{l\neq k}J_{lk}\overrightarrow{n}%
_{l}\overrightarrow{n}_{k}\;.
\end{equation}%
If the non-perturbed spin configuration is described by the set $\left\{ 
\overrightarrow{n}_{l}\right\} $ and the set $\left\{ \overrightarrow{n}%
_{l}+\delta _{il}\delta \overrightarrow{n}_{i}\right\} $ stands for the
perturbed configuration, we obtain for the non-perturbed energy that%
\begin{equation}
E_{0}=-\sum_{l\neq k}J_{lk}\overrightarrow{n}_{l}\overrightarrow{n}_{k}\;.
\end{equation}%
We determine the energy deviation, $\delta E_{i}^{one}$, if we rotate a spin at site $%
i $. If $E(\delta \overrightarrow{n}_{i})$ denotes the energy of the
perturbed configuration then%
\begin{equation}
\delta E_{i}^{one}=E(\delta \overrightarrow{n}_{i})-E_{0}\;.
\end{equation}%
It can be written that 
\begin{align}
E(\delta \overrightarrow{n}_{i})& =-\sum_{l\neq k}J_{lk}\left( 
\overrightarrow{n}_{l}+\delta _{il}\,\delta \overrightarrow{n}_{i}\right)
\left( \overrightarrow{n}_{k}+\delta _{ik}\,\delta \overrightarrow{n}%
_{i}\right) \;,
\end{align}%
i.e.,%
\begin{equation}
\delta E_{i}^{one}= -2\sum_{m\left( \neq i\right) }J_{mi}\overrightarrow{n}%
_{m}\delta \overrightarrow{n}_{i}.  \label{dei}
\end{equation}%
We can determine the energy deviation of
two sites rotation, $\delta E_{ij}$, i.e., we rotate the spins at site $i$
and site $j$ with vector $\delta \overrightarrow{n}_{i}$ and $\delta 
\overrightarrow{n}_{j}$ at the same time, respectively. This energy deviation is calculated as follows,
\begin{equation}
\delta E_{ij}^{two}= E(\delta \overrightarrow{n}_{i},\delta \overrightarrow{n}%
_{j})-\delta E_{i}^{one}-\delta E_{j}^{one} \;,  \label{dH1}
\end{equation}%
which can be written as
\begin{equation}
\delta E_{ij}^{two}=-2J_{ij}\,\delta \overrightarrow{n}%
_{i}\delta \overrightarrow{n}_{j}\;.  \label{dH}
\end{equation}

\subsection{Bilinear-tensorial Heisenberg model}

Here, we assume that there is a tensorial coupling between the magnetic spin moments in the effective Hamiltonian. It can be written that \begin{equation}
\mathcal{H}^{T}=-\sum_{l\neq k}\overrightarrow{n}_{l}%
\mathbf{J}_{lk}\overrightarrow{n}_{k}\;,
\end{equation}%
where $\mathbf{J}_{ij}=\{J_{ij}^{\mu \nu };\mu,\nu\in\{x,y,z\}\}$. We determine the energy deviation, $\delta E_{i}^{one}$, and interaction part of the two-site energy deviation, $\delta E_{ij}^{two}$ as in the scalar model. Every step is the same if we use a tensor coupling instead of scalar, therefore we here only write the final results,
\begin{equation}
\delta E_{i}^{one}= -2\sum_{m\left( \neq i\right) }\overrightarrow{n}_{m}%
\mathbf{J}_{mi}\delta \overrightarrow{n}_{i} \:,
\end{equation}%
and
\begin{equation}
\delta E_{ij}^{two}= E(\delta \overrightarrow{n}_{i},\delta 
\overrightarrow{n}_{j})-\delta E_{i}-\delta E_{j}=-2\delta \overrightarrow{n}%
_{i}\mathbf{J}_{ij}\delta \overrightarrow{n}_{j}\;.  \label{bitens}
\end{equation}
 
\subsection{Biquadratic model}

Here, we express the one- and two-site energy deviations in terms of  a biquadractic effective Hamiltonian. 
\begin{equation}
\mathcal{H}^{Q} =-\sum_{l\neq k}J_{lk}^{\prime }%
\overrightarrow{n}_{l}\overrightarrow{n}_{k}-\sum_{l\neq k}B_{lk}\left( 
\overrightarrow{n}_{l}\overrightarrow{n}_{k}\right) ^{2}\;,  \label{biquad}
\end{equation}%
We can write for the one-site energy deviation that%
\begin{equation}
\delta E_{i}^{one}= -2\sum_{m\left( \neq i\right) }J_{mi}^{\prime }%
\overrightarrow{n}_{m}\delta \overrightarrow{n}_{i}-4\sum_{m\left( \neq
i\right) }B_{mi}\left( \overrightarrow{n}_{m}\overrightarrow{n}_{i}\right)
\left( \overrightarrow{n}_{m}\,\delta \overrightarrow{n}_{i}\right) \;
 \label{onebi}
\end{equation}%
in first order of $\delta \overrightarrow{n}_{i}$.

In case of two-site rotations
\begin{equation}
\delta E_{ij}^{two}=\delta E_{ij}^{QHT}+\delta E_{ij}^{QAT}\;,  \label{dQ}
\end{equation}%
where
\begin{equation}
\delta E_{ij}^{QHT}=-\left[ 2J_{ij}^{\prime }+4B_{ij}\left( \overrightarrow{n}_{i}%
\overrightarrow{n}_{j}\right) \right] \delta \overrightarrow{n}_{i}\delta 
\overrightarrow{n}_{j} \;  \label{dQ}
\end{equation}%
and
\begin{equation}
\delta E_{ij}^{QAT}=-4B_{ij}\left( \overrightarrow{n}_{i}\delta 
\overrightarrow{n}_{j}\right) \left( \overrightarrow{n}_{j}\delta 
\overrightarrow{n}_{i}\right) \;.  \label{dQ}
\end{equation}%
In the collinear limit
\begin{equation}
\delta E_{ij}^{two}=-\left[ 2J_{ij}^{\prime } \pm 4B_{ij} \right] \delta \overrightarrow{n}_{i}\delta 
\overrightarrow{n}_{j}-4B_{ij} \delta n_{i}^{z} \delta n_{j}^{z} \;.  \label{dQz}
\end{equation}
Note in Eq. (\ref{dQz}) that the first term either is a sum of $J_{ij}^{\prime}$ and $B_{ij}$ or the difference between them. The case where $J_{ij}^{\prime}$ and $B_{ij}$ are summed should be used in a calculation with a ferromagnetic configuration, whereas the formula with a minus sign between $J_{ij}^{\prime}$ and $B_{ij}$ should be used for an antiferromagnetic configuration.